\documentclass[11pt]{article}
\usepackage{graphics,epsfig}
\usepackage[]{graphicx}
\usepackage{latexsym}

\begin{document}
\title{Neither Name, Nor Number}
\author{{\sc Federico Holik}}

\maketitle

\begin{center}

\begin{small}
Instituto de Astronom\'{\i}a y F\'{\i}sica del Espacio (IAFE)\\
Casilla de Correo 67, Sucursal 28, 1428 Buenos Aires, Argentina\\

\end{small}
\end{center}

\vspace{1cm}

\begin{abstract}
\noindent

Since its origins, Quantum mechanics has presented problems with the
concept of individuality. It is argued that quantum particles do not
have individuality, and so, one can speak about ``entities without
identity''. On the contrary, we claim that the problem of quantum
non individuality goes deeper, and that one of its most important
features is the fact that there are quantum systems for which
particle number is not well defined. In this work, we continue this
discussion in relation to the problem about the one and the many.

\end{abstract}
\bigskip
\noindent

\newtheorem{theo}{Theorem}[section]
\newtheorem{definition}[theo]{Definition}

\newtheorem{lem}[theo]{Lemma}

\newtheorem{prop}[theo]{Proposition}

\newtheorem{coro}[theo]{Corollary}

\newtheorem{exam}[theo]{Example}

\newtheorem{rema}[theo]{Remark}{\hspace*{4mm}}

\newtheorem{example}[theo]{Example}

\newcommand{\proof}{\noindent {\em Proof:\/}{\hspace*{4mm}}}

\newcommand{\qed}{\hfill$\Box$}

\newcommand{\ninv}{\mathord{\sim}} 

\begin{small}
\centerline{\em Key words:  quasisets, particle number,
quasicardinality, quantum indistinguishability.}
\end{small}

\bibliography{pom}

\newtheorem{axiom}[theo]{Axiom}

\newpage

\section{Introduction}

The concept of individuality in quantum mechanics clashes radically
with its classical counterpart. In classical physics, particles can
be considered as individuals without giving rise to consistency
problems but, in quantum mechanics this is not the case. Problems
arise if one intends to individuate elementary particles. The
responses to this problem range from the claim that there are no
elementary particles at all to the assertion that there are
particles but they are intrinsically indistinguishable (i.e.,
indistinguishable in an ontological sense). Some authors talk about
``entities without identity'' \cite{fyk}. In this work, we claim
that the problem of quantum non individuality is even worse: quantum
non individuality clashes with the concept of number. In this work
we discuss the significance of the superpositions of states with
different particle number in the Fock-space formalism. We continue
to interpret them as systems with an ontologically undefined
particle number. And we conclude that quantum systems not only
suffer the loss of transcendental identity; they also loss the
property of having a definite number. A quantum system is not a
\emph{one}, nor a \emph{many}. But in spite of it, it \emph{is}.

In relation to the problem of quantum indistinguishability, Michael
Readhead and Paul Teller claim in \cite{redtel92} that:

\begin{quotation}``Interpreters of quantum mechanics largely agree that
classical concepts do not apply without alteration or restriction to
quantum objects. In Bohr's formulation this means that one cannot
simultaneously apply complementary concepts, such as position and
momentum, without restriction. In particular, this means that one
cannot attribute classical, well defined trajectories to quantum
systems. But in a more fundamental respect it would seem that
physicists, including Bohr, continue to think of quantum objects
classically as individual things, capable, at least conceptually, of
bearing labels. It is this presumption and its implications which we
need to understand and critically examine.'' M. Readhead and P.
Teller (\cite{redtel92}, p.202)\end{quotation}

When individuality of quanta is studied exhaustively, most
investigations seem to point in the direction that quanta have no
individuality at all (see for example \cite{fyk} for a detailed
analysis). To put it in Schr\"{o}dinger's words:

\begin{quotation}``I mean this: that the elementary particle is not an individual; it
cannot be identified, it lacks `sameness'. The fact is known to
every physicist, but is rarely given any prominence in surveys
readable by nonspecialists. In technical language it is covered by
saying that the particles `obey' a newfangled statistics, either
Einstein-Bose or Fermi-Dirac statistics. [...] The implication, far
from obvious, is that the unsuspected epithet `this' is not quite
properly applicable to, say, an electron, except with caution, in a
restricted sense, and sometimes not at all." E. Schr\"{o}dinger
(\cite{Sch98}, p.197)\end{quotation}

This work should be considered as a tentative to enlarge the problem
posed in the quotation of Readhead and Teller cited above. We claim
that it is usually supposed that quantum systems can be considered
as singular unities, (a quntum system as a ``one"), or collections
of them (a quntum system as a ``many"). In this work we want to
``understand and critically examine" this assumption, thus
continuing the developments in \cite{Domenech-Holik-deRonde}.

In section \ref{s:QinQM}, we discuss the meaning of superpositions
of particle number eigenstates in Fock-space. In \ref{s:Result of a
process} we discuss the assumptions and limitations of the notion of
particles aggregate. Then, in \ref{s:Quai-set}, we review the
approach of quasiset theory to quantum indistinguishability. In
section \ref{s:Modify}, we review the modifications to
Quasiset-theory made in \cite{Domenech-Holik} and in
\ref{s:enlargement} we discuss an enlargement of the Manin's
problem.

\section{Quantity in Quantum Mechanics}\label{s:QinQM}
As is well known, performing a single measurement in a quantum
system does not allow to attribute the result of this measurement to
a property which the system possesses before the measurement is
performed without giving rise to serious problems
\cite{Mittelstaedt}. What is the relationship between this fact and
the quantity of particles in a quantum system? Take for example the
electromagnetic field (with a single frequency for simplicity) in
the following state:
\begin{equation}\label{e:GATOS}
\mid\psi\rangle=\alpha\mid 1\rangle+\beta\mid 2\rangle
\end{equation}
where $\mid 1\rangle$ and $\mid 2\rangle$ are eigenvectors of the
particle number operator with eigenvalues $1$ and $2$ respectively,
and $\alpha$ and $\beta$ are complex numbers which satisfy
$\mid\alpha\mid^{2}+\mid\beta\mid^{2}=1$. If a measurement of the
number of particles of the system is performed, one or two particles
will be detected, with probabilities $\mid\alpha\mid^{2}$ and
$\mid\beta\mid^{2}$ respectively. Any other possibility is excluded.
Suppose that in a single measurement two particles are detected.
What allows us to conclude that the system had two particles before
the measurement was performed? The assertion that the number of
particles is varying in time because particles are being constantly
created and destroyed is also problematic, because it assumes that
at each instant the number of particles is well defined. Only in
case that it is known with certainty that the system is in an
eigenstate of the particle number operator we can say that the
system has a well defined cardinal. There would be no problem too if
it is known with certainty that the system is prepared in an
statistical mixture. In this case, the corresponding density
operator would be:
\begin{equation}
\rho_{m}=\mid\alpha\mid^{2}\mid 1\rangle\langle
1\mid+\mid\beta\mid^{2}\mid 2\rangle\langle 2\mid
\end{equation}
where the subindex ``m'' stands for statistical mixture. But the
density operator corresponding to (\ref{e:GATOS}) is:
\begin{equation}
\rho=(\alpha\mid1\rangle
+\beta\mid2\rangle)(\alpha^{*}\langle1\mid+\beta^{*}\langle2\mid)
\end{equation}
which is the same as:
\begin{equation}\label{e:MGATO}
\rho=\mid\alpha\mid^{2}\mid 1 \rangle\langle
1\mid+\mid\beta\mid^{2}\mid 2\rangle\langle
2\mid+\alpha\beta^{*}\mid 1\rangle\langle2\mid+\alpha^{*}\beta\mid
2\rangle\langle 1\mid
\end{equation}
The presence of interference terms in the last equation implies that
difficulties will appear in stating that, after a single
measurement, the system has the quantity of particles obtained as
the result of the measurement. In this case, the incapability of
knowing the particle number would not come from our ignorance about
the system, but from the fact that in this state, the particle
number is not even well defined. We see thus, that if superpositions
of particle number are alllowed, we are faced with an
\emph{indeterminancy} in the particle number. An indeterminancy of
the same kind of that wchich appears in position or spin. But this
time, the ``property'' affected is very special; \emph{it is just
the property linked to the dicothomy of being one or many}.

\section{How many?}\label{s:Result of a process}
Let us discuss the origin of the concept of ``particle number''. We
start posing the questions: In which sense do we talk about quantum
systems composed, for example, of a single photon? How do we decide
if the field is in a single photon state or not? What do we mean
when we use the words ``single photon''? We could search for a clue
to answer these questions in our laboratory experience, i.e., making
measurements. In experiments, we often use a picture which allows us
to speak about the photon as a particle (and so, as an individual).
In a similar way, and always in relation to experiments, we talk
about the other particles (electrons, protons, etc.). But in a
deeper analysis we find that this supposed ``particle behaviour" of
quantum systems is hardly compatible with its classical counterpart,
and though experiments seem to suggest an idea of individuality, it
is well established that this does not enable us to consider
particles as individuals. Individuality is not compatible with the
formal structure of quantum mechanics \cite{Domenech-Holik-deRonde}
and so elementary particles cannot be considered as individuals, as
E. Schr\"{o}dinger pointed out in the early days of the theory
\cite{Sch98}. In spite of these difficulties, we continue speaking
about photons, electrons, etc., using a jargon which has a lot of
points in common with classical physics, source of conceptual
confusion.

But it is just this interpretation, which presuposes the concept of
`particle' (or quantum object, to put it in more general terms)
which lies at the heart of the notion of `particles aggregate' (and
so, a \emph{defined} particle number, or a \emph{defined} number of
objects). There are definite experimental arrangements which force
the appearance of particle characteristics as a final result of a
single process. Experiments are designed to find out which is the
particle number, but as we have already mentioned in \ref{s:QinQM},
this does not imply that the resulting number is a property that
pertains to the system under study. On the contrary, it refers to
the definite {\it process} which takes place in each measurement. We
are not allowed to consider the system as an aggregate of
individuals as if they were classical objects, simply because the
notion of ``object" is incompatible with the formal structure of the
theory \cite{Domenech-Holik-deRonde}. The fall of the notion of
``object", causes the fall of the notion of ``objects aggregate".

We claim that superpositions in particle number discussed in
\ref{s:QinQM} are a direct expression of the fall of the concept of
``objects aggregate". We know that it is possible to assign to some
quantum systems an associated number, take for example the electrons
of a Litium atom, or single photon states. But particle number
superpositions show that in general, it is not true that a definite
number can be always assigned in a consistent manner. We are
accustomed in our classical experience to assign always a definite
number to the set of things that we are studying. But when we are
faced to a quantum system, we are forced to abandon that habit.

\section{Quasi-set theory and indistinguishability}\label{s:Quai-set}
In the standard approach to quantum indistinguishability, particles
are labeled as if they were individuals, and then
indistinguishability is recovered via simmetrization postulates
\cite{redtel92}. This is a variant of what in \cite{Why quasisets}
was called the Weyl's strategy. Many authors pointed out the
importance of developping alternative ways to describe quantum
indistinguishability, reproducing the results obtained by standard
techniques, but assumming in every step of the deduction that
elementary particles of the same class are intrinsically
indistinguishable from the beginning (see, for example, \cite{fyk},
\cite{Why quasisets} and \cite{Heinz Post}), without making appeal
to Weyl's strategy variants. Another claim is that quantum mechanics
does not possess its own language, but it uses a portion of
functional analysis which is itself based on set theory, and thus
finally related to classical experience. This statement was posed by
Y. Manin \cite{Manin2}, the Russian mathematician who suggested that
standard set theories (as Zermelo-Fraenkel, $ZF$) are influenced by
every day experience, and so it would be interesting to search for
set theories which inspire its concepts in the quantum domain. This
is known as Manin's problem \cite{Manin}. In this spirit, and
looking for a solution to Manin's problem a quasiset theory
($\mathcal{Q}$ in the following) was developed \cite{Why quasisets},
\cite{Un estudio}.

Quasiset theory seems to be adequate to represent as ``sets'' of
some kind (quasisets) the collections of truly indistinguishable
entities. This aim is reached in $\mathcal{Q}$ because equality is
not a primitive concept, and there exist certain kinds of
\textit{urelemente} (m-atoms) for which only an indistinguishability
relationship applies. So, in $\mathcal{Q}$, non individuality is
incorporated by proposing the existence of entities for which it has
no sense to assert that they are identical to themselves or
different from others of the same class.

$\mathcal{Q}$ contains a copy of Zermelo-Fraenkel set theory plus
Urelemente ($ZFU$). These Urelemente are called M-atoms. This
feature divides the theory in two parts. One region involves only
the elements of $ZFU$, and the other one contains quasisets whose
elements can be truly indistinguishable entities. Quasisets
containing only indistinguishable elements are called ``pure
quasisets''. The $ZFU$ copy of quasisets is called ``the classical
part of the theory '' in \cite{Un estudio}. Indistinguishability is
modeled in this theory using a primitive binary relation $\equiv$
(indistinguishability) and a new class of atoms, called m-atoms,
which express the existence of quanta in the theory \cite{Un
estudio}. So, in the frame of $\mathcal{Q}$, when we speak of
m-atoms of the same class, the only thing that we can assert about
them is that they are indistinguishable, and nothing else makes
sense, for expressions like $x=y$ are not well formed formulas. This
is to say that we cannot make assertions about their identity, i.e.,
it has no sense to say that an m-atom is equal or different of other
m-atom of the same class. It is important to remark that in
$\mathcal{Q}$, indistinguishability does not imply identity, and so
it is possible that even being indistinguishable, two m-atoms belong
to different quasisets, thus avoiding the collapse of
indistinguishability in classical identity \cite{Un estudio}.

$\mathcal{Q}$ is constructed in such a way that allows the existence
of collections of truly indistinguishable objects, and thereof it is
impossible to label the elements of pure quasisets. For this reason,
the construction used to assign cardinals to sets of standard $ZFU$
theories cannot be applied any more. But even if electrons are
indistinguishable (in an ontological sense), every physicist knows
that it makes sense to assert that, for example, a Litium atom has
three electrons. It is for that reason that $\mathcal{Q}$ should
allow quasisets to have some kind of associated cardinal. In
$\mathcal{Q}$ this is solved postulating that a cardinal number is
assigned to every quasiset (remember that there is a copy of $ZFU$
in $\mathcal{Q}$). Some other properties of the standard cardinal
are postulated too. This rule for the assignment of cardinals uses a
unary symbol $qc()$ as a primitive concept. So in $\mathcal{Q}$, the
quasicardinal is a primitive concept alike $ZF$, in which the
property that to every set corresponds a single cardinal number can
be derived from the axioms \cite{Halmos}.

An important theorem of $\mathcal{Q}$ is related to the
unobservability of permutations:

{\sf [Unobservability of Permutations]}\label{unobservabilty} Let
$x$ be a finite quasi-set such that $x$ does not contain all
indistinguishable from $z$, where $z$ is an $m$-atom such that $z
\in x$. If $w \equiv z$ and $w \notin x$, then there exists $w'$
such that
 $$(x - z') \cup w' \equiv x$$

It is the assertion in the language of $\mathcal{Q}$ that
permutation of indistinguishable quanta cannot imply any observable
effect, or put in words of Penrose:

\begin{quotation}
``[a]ccording to quantum mechanics, any two electrons must
necessarily be completely identical [in the physicist's jargon, that
is, indistinguishable], and the same holds for any two protons and
for any two particles whatever, of any particular kind. This is not
merely to say that there is no way of telling the particles apart;
the statement is considerably stronger than that. If an electron in
a person's brain were to be exchanged with an electron in a brick,
then the state of the system would be \textit{exactly the same
state} as it was before, not merely indistinguishable from it! The
same holds for protons and for any other kind of particle, and for
the whole atoms, molecules, etc. If the entire material content of a
person were to be exchanged with the corresponding particles in the
bricks of his house then, in a strong sense, nothing would be
happened whatsoever. What distinguishes the person from his house is
the \textit{pattern} of how his constituents are  arranged, not the
individuality of the constituents themselves'' \cite[p.\ 32]{pen89}.
\end{quotation}

In the next section, we go back to the problem of an undefined
particle number and we relate it with $\mathcal{Q}$.

\section{Modifications to the theory of Quasi-sets}\label{s:Modify}
The way in which the quasicardinal is introduced in $\mathcal{Q}$
implies that every quasiset has an associated cardinal, i.e., every
quasiset has a well defined number of elements. But the idea that an
aggregate of entities must necessarily have an associated number
which represents the number of entities is based in our every day
experience. As we have mentioned in section \ref{s:QinQM}, there are
quantum systems to which it is not allowed to assign a number of
particles in a consistent manner.

Taking into account these considerations, it is worth asking: is it
possible to represent a system prepared in the state (\ref{e:MGATO})
in the frame of quasiset theory? Which place would correspond to a
system like (\ref{e:MGATO}) in that theory? If such system could be
represented as a quasiset, then it should have an associated
quasicardinal, for every quasiset has it. But this does not seem to
be proper, considering what we have disscussed in section
\ref{s:QinQM}. It follows that it does not appear reasonable to
assign a quasicardinal to every quasiset if quasiset theory has to
include all bosonic and fermionic systems (in all their possible
many particle states). Therefore, a system in the state
(\ref{e:MGATO}) cannot be included in $\mathcal{Q}$ as a quasiset.
Yet, it would be interesting to study the possibility of including
systems in those states (such as (\ref{e:MGATO})) in the formalism.

A possible way out would be the introduction of a Fock-space, but
constructed using the non classical part of $\mathcal{Q}$. This
option has been considered in \cite{Domenech-Holik-Krause}. In that
paper, we reformulate Fock-space quantum mechanics using
$\mathcal{Q}$. Due to the theorem of unobservability of permutations
(see last section of this work), we avoid the use of the
\emph{labeled-tensor-product-Hilbert-space formalism} (LTPHSF), as
called by Redhead and Teller \cite{redtel91}, \cite{redtel92}. So we
can express states such as (\ref{e:MGATO}) as superpositions of
different particle number in this novel space.

Another possibility is to reformulate $\mathcal{Q}$ in such a way
that the quasicardinal is not to be taken as a primitive concept,
but as a derived one, turning it into a property that some quasisets
have and some others do not (in analogy with the property ``being a
prime number'' of the integers). Those quasisets for which the
property of having a quasicardinal is not satisfied, would be
suitable to represent quantum systems with particle number not
defined (such as equation (\ref{e:MGATO})). This property would also
fit well with the position that asserts that particle interpretation
is not adequate in, for example, quantum electrodynamics. With such
a modification of $\mathcal{Q}$, for example, a field (in any state)
could always be represented as a quasiset, avoiding the necessity of
regarding the field as a collection of classical ``things''. On the
contrary, the field would be described by a quasiset which has a
defined quasicardinal only in special cases, but not in general.
Thus a field would be represented as something which genuinely is
nor a one nor a many. And for that reason this quasiset could not be
interpreted so simply as a collection of particles (because it seems
reasonable to assume that a collection of particles,
indistinguishable or not, must always have a well defined particle
number).

We have followed the idea of modifying $\mathcal{Q}$ in
\cite{Domenech-Holik}. In that work, we have searched for a
$\mathcal{Q}$-like theory in which the quasicardinal is not taken as
a primitive concept alike $\mathcal{Q}$. We showed that it is
possible to develop a theory about collections of indistinguishable
entities in which quasicardinal is not a primitive concept. We
define a singleton which allows us to extract ``just one element"
 form a given quasiset $X$, and obtain a subquasiset $X^-$ such that:

             $$X \supset X^-$$

\noindent It is important to remark that it has no sense to ask
which is the extracted element, because this query is not defined in
a theory without identity. It could happen that $X^-=\emptyset$ or
not. If $X^-\neq_E\emptyset$ it follows that we can make the same
operation again, and obtain $X^{--}$. Then we have:

$$X \supset X^{-} \supset X^{--} $$

Going ahead with this process, it could be the case that this chain
of inclusions stopped (in case the last quasiset so obtained be the
empty quasiset), or that it has no end. So we could conceive two
qualitatively different situations:

Situation $1$: $$X \supset X^{-} \supset X^{--} \supset X^{---}
\supset \cdots \cdots$$ (the inclusions chain continues
indefinitely)

Situation $2$: $$X \supset X^{-} \supset X^{--} \supset X^{---}
\supset \cdots \cdots \supset \emptyset$$ (the inclusion chain ends
in the empty quasiset).

We call \emph{descendent chains} to the chains of inclusions which
appear in situations $1$ and $2$. In order to grant that one of
these situations holds for every quasiset we postulate the existence
of descendent chains. This is done by first translating the concept
of descendent chain to first order language:

\begin{definition}\label{e:cdx}
 $$CD_X(\gamma)\longleftrightarrow$$ $$( \gamma \in \wp (\wp
 (X)) \wedge X \in\gamma  \wedge \forall z \forall y (z\in \gamma
 \wedge y\in \gamma\wedge z\neq_{E} y\longrightarrow(z\supset y
 \vee y\supset z))$$
 $$\wedge\forall z(z\in \gamma\wedge
 z\neq_{E}\emptyset\longrightarrow\exists y(y\in\gamma\wedge
 DD_z(y)\wedge\forall w(w\in\gamma\wedge DD_z(w)\longrightarrow
 w=_E y))))$$\
$CD_X(\gamma)$ is read as: $\gamma$ is a descendant chain of $X$.
\end{definition}

In the construction shown in \cite{Domenech-Holik} we have
reobtained that every (finite) quasiset has a well defined
quasicardinal. But in that construction, we can assert that a
quasiset has a definite quasicardinal only if it satisfies the
definition of finiteness, and nothing can be said about the
quasisets which do not satisfy the definition. Though not
necesarilly useful for the problem of the undefined particle number,
the axiomatic variant exposed in \cite{Domenech-Holik} shows
explicitly that in a theory about collections of indistinguishable
entities, the quasicardinal needs not be necessarily taken as a
primitive concept. This result encourages the research of more
complex axiomatic formulations, able to incorporate the quantum
systems with undefined particle number as sets of some kind, thus
enriching Manin's problem.

\section{A new turn on Manin's problem}\label{s:enlargement}
In this article, we have disscussed that quantum systems not only
seem to lack individuality, but moreover, they seem to be undefinite
in their cardinality. What is the meaning of this assertion?
Perhaps, we have to abandon the supposition that we are dealing with
something like ``entities without identity", in the sense that we
cannot consider a quantum system as a ``one" or a ``collection of
ones". Perhaps, besides the ontological presupposition of
individuality, we have to abandon the presupposition of number: so,
we would have \emph{neither name, nor number}. In this sense, the
suggestions of Manin:

\begin{quote}
``We should consider the possibilities of developing a totally new
language to speak about infinity.\footnote{Set theory is also known
as the theory of the `infinite'.} Classical cri\-tics of Can\-tor
(Brou\-wer \textit{et al.}) argued that, say, the general choice
axiom is an illicit extrapolation of the finite case.

   I would like to point out that this is rather an
extrapolation of common-place physics, where we can distinguish
things, count them, put them in some order, etc. New quantum physics
has shown us models of entities with quite different behavior. Even
`sets' of photons in a looking-glass box, or of electrons in a
nickel piece are much less Cantorian than the `set' of grains of
sand. In general, a highly probabilistic `physical infinity' looks
considerably more complicated and interesting than a plain infinity
of `things'.''
\end{quote}

\noindent should be enlarged in order to include the question of
undefined number. It would be very interesting to search for a ``set
theory" inspired on quantum phenomena, which could take into account
the problem of undefined cardinality (perhaps, the term ``set" would
not be appropriate any more in a theory like this). It is quite
possible that the development of a theory like this one, would yield
a new way to approach the problem of quantum separability. Perhaps,
when we speak of a quantum system composed of two subsystems we are
thinking of it \emph{as a collection of things}, and this could be
linked to the problems which emerge in quantum separability, because
we have a wrong way to speak about it.

$\mathcal{Q}$ gives an answer to Manin's problem, in the sense that
it is a ``set theory" concerning collections of truly
indistinguishable objects. But in order to solve the extended
version of Manin's problem, further formal developments needs to be
achieved. The axiomatic variant presented in \cite{Domenech-Holik}
shows that in a theory like $\mathcal{Q}$, the quantity of elements
needs not to be a primitive concept. This result encourages the
search for a theory in which it is impossible to assign a
quasicardinal to certain quasisets in a consistent manner, thus
allowing to describe what it seems to happen with some quantum
systems, in which non individuality expresses itself in the fact
that particle number is not defined, besides ontological
indistinguishability.

\vskip1truecm

\noindent {\bf Aknowledgements} \noindent The author whish to thank
Graciela Domenech and Christian de Ronde for useful discussions.
\noindent This work was partially supported by the following grants:
PICT 04-17687 (ANPCyT), PIP N$^o$ 6461/05 (CONICET), UBACyT N$^o$
X204.

\end{document}